\documentclass[pra,twocolumn,english,superscriptaddress,floatfix]{revtex4-2}
\usepackage[utf8]{inputenc}
\usepackage{graphicx}
\usepackage{dcolumn}
\usepackage{bm}
\usepackage{amsmath,amssymb}

\makeatletter
\newcommand{\lesscong}{\mathrel{\gl@over{<}{\cong}}}
\newcommand{\lesssimeq}{\mathrel{\gl@over{<}{\simeq}}}
\newcommand{\gtrcong}{\mathrel{\gl@over{>}{\cong}}}
\newcommand{\gtrsimeq}{\mathrel{\gl@over{>}{\simeq}}}
\newcommand{\gl@over}[3][1pt]{%
  \vcenter{\m@th\offinterlineskip\ialign{%
    \hfil$##$\hfil\cr #2\cr \noalign{\vskip#1} #3\cr
  }}%
}
\usepackage{mathtools}
\usepackage{xcolor}
\usepackage{braket}
\usepackage{url}
\usepackage{bbold}
\usepackage{indentfirst}
\usepackage{braket}
\usepackage{qcircuit}
\usepackage{float}

\def\bracket#1#2{{\langle#1|#2\rangle}}

\begin{document}
\preprint{APS/123-QED}

\title{Measure and Forget Dynamics in Random Circuits}

\author{Yucheng He}
\email{yucheng.he@usc.edu}
\affiliation{Ming Hsieh Department of Electrical and Computer Engineering, University of Southern California, Los Angeles, California, USA}
 
\author{Todd A. Brun}\email{tbrun@usc.edu}
\affiliation{Ming Hsieh Department of Electrical and Computer Engineering, University of Southern California, Los Angeles, California, USA}%



\date{\today}

\begin{abstract}
Unrecorded or ``forgetful'' measurements---physically, a local dephasing channel---can arise naturally as syndrome-measurement errors in fault-tolerant protocols and as uncontrolled decoherence in open systems. We numerically study random Clifford circuits in which recorded projective $Z$ measurements compete with unrecorded (forgotten) $Z$ measurements, focusing on the dynamical, finite-depth regime relevant to decoding and quantum memory. In the forget-only limit the global entropy density thermalizes at a rate independent of system size, and the forgetting rate at which thermalization sets in decays with circuit depth as a power law $p_f^{\ast}\sim d^{v}$, consistent with the critical-depth scaling inversely with the noise rate found for related noisy-circuit quantities\cite{CMISpreading,NoisyEncoding}. With recorded measurements present, the mutual information $I(A{:}B)$---which quantifies the recoverable correlation between the two halves---develops a finite-depth peak whose height grows sub-extensively with $N$: its effective scaling exponent $\alpha(p_f)$ decreases monotonically, so that forgetting drives the recoverable correlation from near-volume-law toward area-law scaling. We map the recoverability landscape in the $(p_m,p_f)$ plane and show that forgetting destroys the measurement-induced purification transition. These results complement earlier stat-mech analyses of noisy monitored circuits~\cite{ZhiLi,DiasNoise} and quantify the competition between measurement and forgetting relevant to quantum error correction.
\end{abstract}

\maketitle

\section{Introduction}
\label{secI}

Random quantum circuits, a fascinating and rapidly evolving area of research at the intersection of quantum complexity theory \cite{FOCS,Efficient,Bouland2019,Noh2020efficientclassical} and condensed matter physics \cite{Matteo,ChaomingCriticality,ChaomingPhysRevBtheory}, offer insights into the behavior and capabilities of quantum systems. These circuits, composed of a sequence of randomly chosen local unitary gates, are versatile tools for exploring a wide range of quantum phenomena, from quantum entanglement \cite{Zhang2022,PhysRevX.10.031066} and chaos \cite{PhysRevX.10.031066} to quantum error correction \cite{SoonwonChoi,YoshidaCode,Ruihua,Dehghani2023NNDecoder,LeeYoshida2024RandomlyMonitored} and measurement-induced entanglement transitions \cite{YimuBao,MGullans}. We note that these monitored-dynamics transitions are distinct from conventional ground-state quantum phase transitions.

Starting from this setup of random quantum circuits, there have been a number of studies of the effects of random local measurements among the layers of gates, leading to a kind of dynamics called measurement-induced phase transition (MIPT) \cite{YaodongZeno,SkinnerRuhmanNahum,ChanNandkishore,ChaomingPhysRevBtheory,YimuBao,YDLiDomainWall}. In the context of bipartite isolated quantum systems, the entanglement entropy gradually expands over time, ultimately saturating to a value proportional to the system's size \cite{NahumGrowth}. Introducing projective measurements at random locations acts as a mitigating force, reducing the entanglement and suppressing the entropy growth. This interplay gives rise to a notable phenomenon: as the measurement rate increases within a non-integrable systems, a phase transition in the entanglement entropy becomes apparent. In the long-time limit, the entropy shifts from being determined by the size of the subsystem (known as the ``volume law'' phase) to being determined by the boundary area (the ``area law'' phase). This kind of phase transition has been widely discussed theoretically \cite{YaodongZeno,YDLiDomainWall,ZhangZhang2022NoiseTransition,Liu2024NoisyHybridPRL,Google2023MIPTExperiment} and verified experimentally \cite{Noel2022,Koh2023}.

The possibility that the dynamics of random quantum circuits might be useful in developing quantum devices or coding protocols---such as low-depth random codes for quantum memory \cite{GullansLowDepth,BrownFawzi}---provides motivation to study their long-time behavior, and especially to understand what happens as the system ultimately forgets the quantum information \cite{Fidkowski2021howdynamicalquantum}.



In this paper we study random Clifford circuits in which recorded projective $Z$ measurements compete with unrecorded (``forgotten'') $Z$ measurements, the latter acting as a local dephasing channel; for a recent overview of noisy monitored circuits see Ref.~\cite{NoisyMonitoredReview}. Two neighboring regimes have been analyzed via a classical statistical-mechanics mapping. For noise alone (no recorded measurement), Li \emph{et al.}~\cite{ZhiLi} found a size-independent thermalization rate, and that bipartite entanglement measures rise to a peak, obeying an area law; adding recorded measurement, Dias \emph{et al.}~\cite{DiasNoise} obtained a steady-state phase diagram and showed that noise smears out the measurement-induced phase transition into a simple crossover. We study the same measurement/forgetting competition in the \emph{dynamical, finite-depth} regime most relevant to decoding and quantum memory. In the forget-only limit we recover a size-independent thermalization rate, now with a power-law dependence of the thermalization ``turning point'' on circuit depth (Sec.~\ref{forgetinduced}). With recorded measurement also present, we characterize how forgetting degrades the mutual information $I(A{:}B)$ (our proxy for recoverability) and map the recoverability landscape (Secs.~\ref{competition} and \ref{appcode}). In particular, the peak mutual information grows sub-extensively, $I_{\rm peak}\sim N^{\alpha(p_f)}$ with $\alpha\in(0,1)$---intermediate between the area law of the noise-only case and a volume law---and we show that forgetting destroys the measurement-induced purification transition (Sec.~\ref{purification}).

\section{Forgetting Induced Dynamics}
\label{forgetinduced}

\subsection{The Forgetting Process}
\label{forget}

In quantum information processing, forgetting a measurement outcome can be a kind of error in an unreliable measurement, such as a ``syndrome measurement error'' in fault-tolerant quantum computing. On the theoretical physics side, natural systems do not have a reliable recorder for measurement results, which can quickly become inaccessible; so studying the forgetting process can help us understand dynamics outside of an artificial system. Moreover, much recent work focuses on measurement-induced entanglement transitions, but the fate of volume-law entanglement when outcomes are discarded remains largely unexplored and could offer valuable insights into the underlying dynamics.

Usually, when we talk about measurement, we include recording the measurement outcome. For example, for a pure state $\ket{\psi}=\alpha\ket{0}+\beta\ket{1}$, 
\begin{equation}
\rho=\ket{\psi}\bra{\psi} =
\begin{pmatrix}
|\alpha|^2&\alpha\beta^* \\ \alpha^*\beta&|\beta|^2
\end{pmatrix}
\end{equation}
the whole process can be represented
\begin{equation}
\rho \xrightarrow{\text{Measure}}
\begin{cases}
\rho_0=\frac{P_0\rho P_0}{p_0} =
\begin{pmatrix}
1&0 \\ 0&0
\end{pmatrix}, & \text{result is 0}, \\
\rho_1=\frac{P_1\rho P_1}{p_1} = 
\begin{pmatrix}
0&0 \\ 0&1
\end{pmatrix}, & \text{result is 1},
\end{cases}
\end{equation}
where $P_{0,1}$ are the projectors $P_0=\ket0\bra0$ and $P_1=\ket1\bra1$, and $p_0 = |\bracket{0}{\psi}|^2$ and $p_1 = |\bracket{1}{\psi}|^2$ are the probabilities for results 0 and 1, respectively.

Now suppose that we measure in the standard basis but do not record the measurement outcome---that is, we forget the result. For the pure state $\ket\psi$, if we measure and then forget the measurement outcome, the state will be
\begin{equation}
\rho\rightarrow
\begin{pmatrix}
|\alpha|^2&0 \\ 0&|\beta|^2
\end{pmatrix},
\end{equation}
which is a mixed state, equivalent to a classical ensemble. This process is equivalent to a ``complete dephasing channel''~\cite{NielsenChuang}. It is the special case of the general dephasing channel
\begin{equation}
\rho \rightarrow (1-p)\rho + pZ\rho Z
\end{equation}
when the probability $p=1/2$. In the rest of the paper, we will refer to the ``measure-and-recording'' process as ``measurement'' and the ``measure-and-forget'' process as ``forget.''

As a quantum channel, the forget process is exactly the complete $Z$-dephasing channel above, applied independently at each site with probability $p_f$. We model this as a heralded process, with dephasing events applied randomly on the qubits with an overall rate $p_f$. At low rates, the results would probably be qualitatively similar for general unheralded dephasing noise but not quantitatively the same. At high rates $p_f$ both cases would lead to the suppression of recoverability found in Sec.~\ref{appcode}, but we would expect the unheralded case to be worse.

As both measurement and forget lead to changes in the system's uncertainty, we use von~Neumann entropy to characterize it:
\begin{equation}
S(\rho)=-\text{Tr}(\rho \log \rho),
\end{equation}
where $\rho$ is the density matrix of the whole system. The state after the measurement process will be in a pure state if it was initially pure; and if it was in a mixed state its von~Neumann entropy will generally go down, and never increase. After the forget process, a pure state will generally become mixed, unless the initial state was already in the measurement basis; and a mixed state will generally have its entropy go up, and never decrease.

\subsection{Random Clifford Model and Entropy Growth}

\begin{figure*}[bbpt]
    \centering
    \includegraphics[scale=0.23]{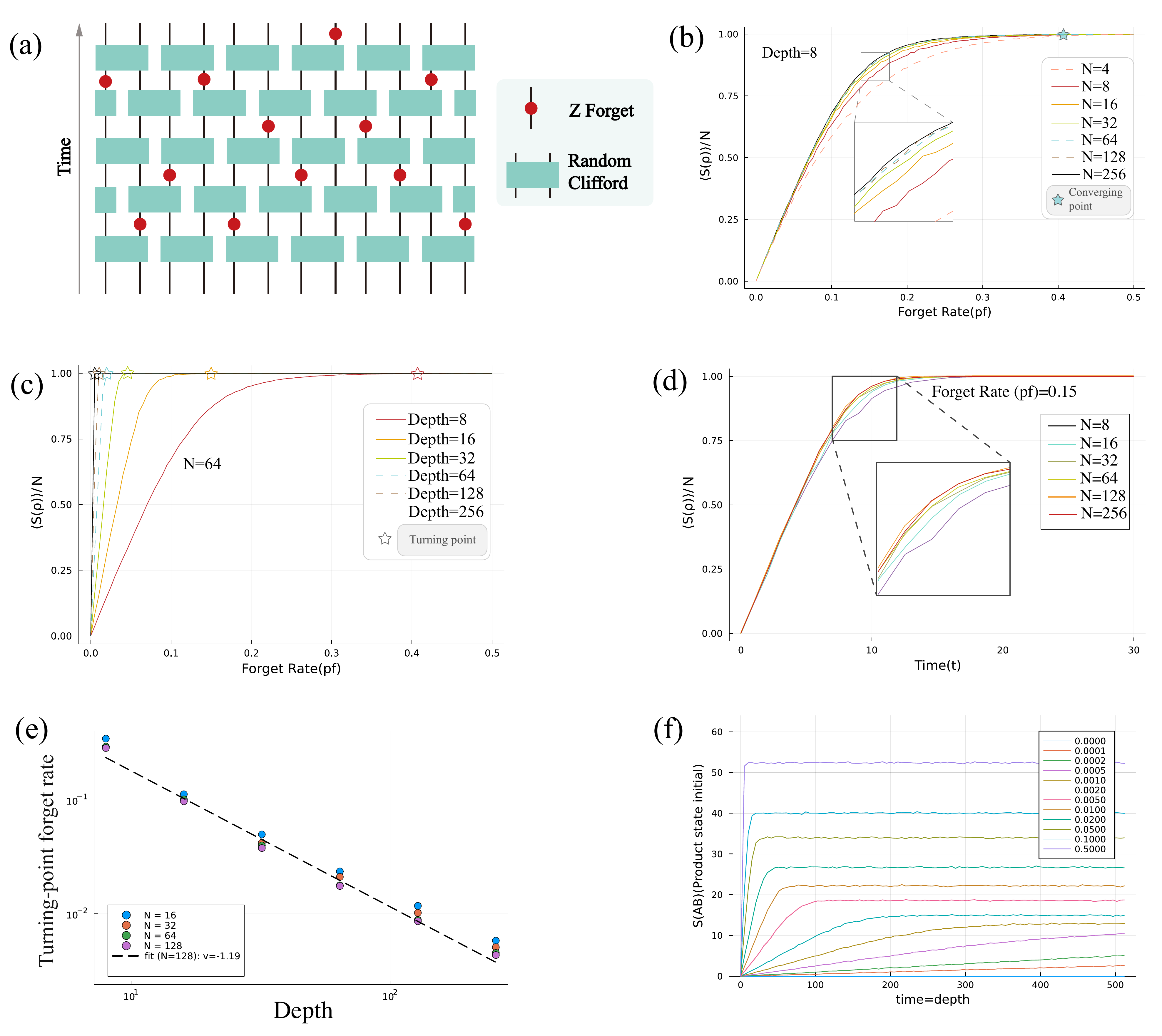}
    \caption{(a) A configuration of local two-qubit unitaries is uniformly drawn from the Clifford group and arranged in a 1D brickwork pattern, with periodic boundary conditions. Between each layer of gates, ``$Z$ forgetting'' happens with probability $p_f$ at each site. The initial state is pure. (b) Entropy changes as a function of forgetting rate for a fixed circuit depth 8. The system shows a local thermalization rate that is independent of system size $N$. (c) Entropy changes as a function of forgetting rate for a fixed system size $N=64$. The turning point where the system first reaches the upper bound shifts toward smaller $p_f$ with increasing circuit depth. (d) Plotting $S(\rho)/N$ as a function of circuit depth (time) for a fixed forgetting rate, the curves for different system sizes are essentially the same. (e) The forgetting rate at the turning point as a function of circuit depth, for several system sizes $N=16,32,64,128$ (log--log). The data collapse onto a single power law $p_f^{\ast}\sim\alpha\,d^{v}$, independent of $N$ (the exponents for $N=64$ and $N=128$ agree to within $0.001$ at fixed threshold). We find an effective exponent $v\approx-1.2$, with a systematic uncertainty $\sim\pm0.08$ set by the choice of saturation threshold; it should be read as an effective finite-size exponent rather than a universal critical exponent, and is consistent with the critical-depth $\propto 1/\mathrm{noise}$ scaling reported for related noisy-circuit quantities~\cite{CMISpreading,NoisyEncoding}. (f) Global von~Neumann entropy $S(\rho)$ versus depth for several forgetting rates $p_f$ at $N=64$: increasing $p_f$ raises the saturation level and shortens the thermalization time.}
    \label{Forget_only}
\end{figure*}

It is now well understood that measurement-induced phase transitions occur in ensembles of quantum trajectories \cite{MGullans}. Such transitions typically arise when there is a balance between unitary dynamics, which increases entanglement within the system, and measurement processes, which reduce entanglement. To explore these entanglement transitions, we study the ``random Clifford'' model (see Fig.~\ref{Forget_only}(a)).

The model consists of a circuit resembling a ``brickwork'' pattern, where random two-site unitary operations are uniformly selected from the Clifford group. This circuit acts on a linear chain of $N$ qubits with periodic boundary conditions. After each layer, every qubit in the circuit is measured in the $Z$ basis with a fixed probability $p_f$, and the outcome is then forgotten, which in turn changes the entropy of the system. This is a specific type of stabilizer circuit, which by the Gottesman--Knill theorem can be simulated on a classical computer in a time that grows polynomially with $N$~\cite{AaronsonGottesman}. It is therefore possible to perform a finite-size scaling analysis even for a large number of qubits (e.g., thousands). All simulations in this work were performed with the {\sc QuantumClifford.jl} package~\cite{QuantumCliffordJl}.

The initial state can take the form of any arbitrary pure stabilizer state. In Fig.~\ref{Forget_only}(b), we have graphed the entropy density $S(\rho)/N$ (the von~Neumann entropy of the whole system per qubit, also called the entropy per site) against the forgetting rate $p_f$. The graph shows an average over 200 realizations. Notably, for a constant depth of 8, the curve reaches its saturation (plateau) value at $p_f\approx0.415$, regardless of the magnitude of $N$; for $N\gtrsim32$ the curves for different $N$ are nearly indistinguishable. This indicates that the entropy density $S(\rho)/N$ relaxes toward its steady-state value at a rate that does not depend on the system size---a behavior we refer to as a \emph{local thermalization rate}.

In Fig.~\ref{Forget_only}(c), the entropy density is displayed for a fixed system size $N=64$ for a variety of depths, plotted against the forgetting rate. The ``turning point'' refers to the point at which this function first reaches its upper bound. 

Fig.~\ref{Forget_only}(d) provides the clearest evidence for this: when the entropy density $S(\rho)/N$ is plotted against time (depth) at a fixed forgetting rate, the curves for different system sizes $N$ collapse onto one another. This is expected from physical arguments: since forgetting acts independently at each site with probability $p_f$ per layer, and the gates are local, entropy is produced at an intensive, per-site rate; the relaxation of the entropy density therefore obeys local dynamics and is independent of the total number of qubits. We note that while both this collapse and the size-independent turning-point scaling discussed below are fully consistent with local thermalization, a complete demonstration would also examine local observables, reduced density matrices of subsystems, or correlation lengths; we leave such analysis to future work.

As the circuit depth increases, the forgetting rate at the turning point rapidly approaches zero. This is shown in Fig.~\ref{Forget_only}(e), where we plot the turning-point forgetting rate versus circuit depth for several system sizes on a log--log scale. The results for $N=16,32,64,128$ collapse onto a single power law $p_f^{\ast}\sim\alpha\,d^{v}$, confirming that the relation is independent of system size; in particular, the fitted exponents for $N=64$ and $N=128$ coincide to within $0.001$ at a fixed saturation threshold. We extract an effective exponent $v\approx-1.2$, with a systematic uncertainty of about $\pm0.08$ arising from the threshold used to define the turning point. We emphasize that this is an effective, finite-size exponent, rather than a universal critical exponent. Its value is consistent with the critical-depth $\propto 1/\mathrm{noise}$ scaling established for related quantities in noisy random circuits---e.g.,\ the divergence depth of the conditional mutual information~\cite{CMISpreading} and the depth threshold for information recovery in noisy encoding circuits~\cite{NoisyEncoding}---and our forget-only result gives a complementary, entropy-based realization of the same scaling family.

\section{Competition Between Measurement and Forgetting}
\label{competition}

\begin{figure}[hpbt]
\centering
\includegraphics[scale=0.37]{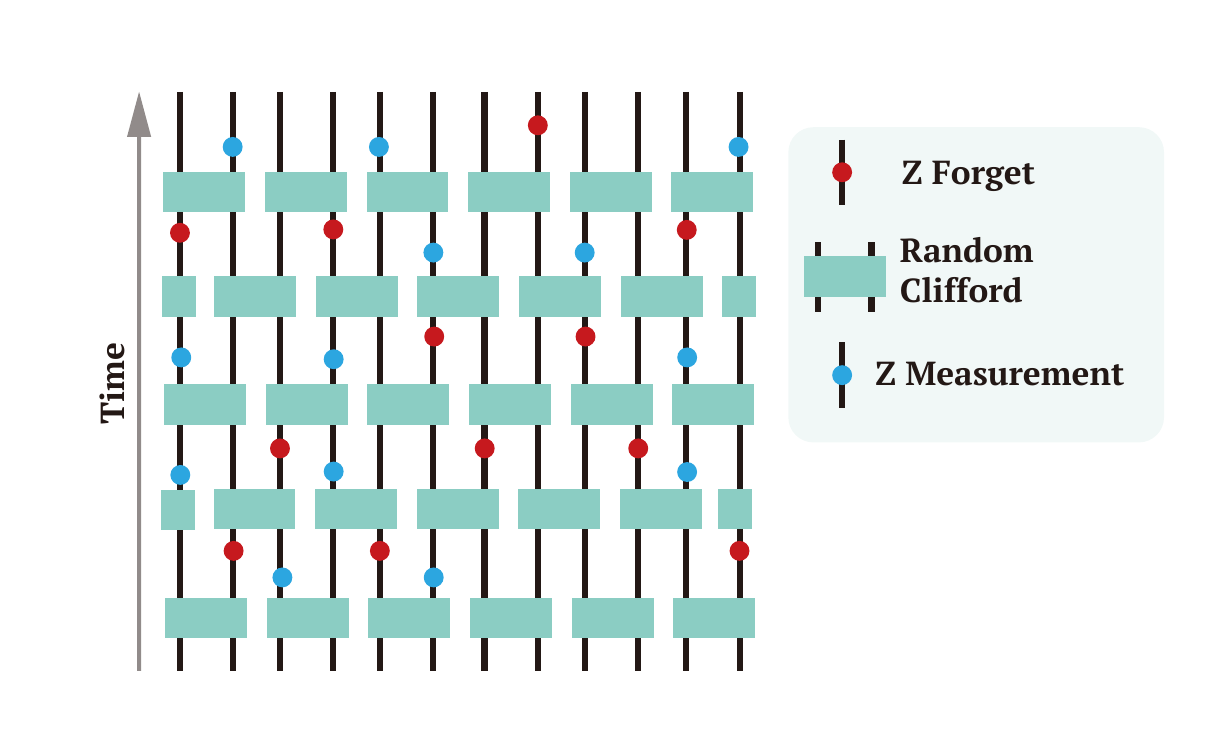}
\caption{The circuit employed to explore the interplay between the measurement and forget processes. Between each ``brickwork'' layer of random Clifford gates, we introduce supplementary strata of random ``$Z$ Measurements'' and ``$Z$ Forgets.'' The boundary conditions are periodic.}
\label{competitionfig}
\end{figure}

Beyond studying forget-only dynamics, it is even more intriguing to explore the competition between the measurement and the forgetting processes. 
The system depicted in Fig.~\ref{competitionfig} comprises three essential components, each with distinct effects: (1) local $Z$ measurement at random sites, which reduces the von~Neumann entropy and also destroys entanglement; (2) random Clifford gates, which maintain entropy and can increase entanglement; and (3) local $Z$ forgetting at random sites, which increases entropy and also destroys entanglement. When combined, these elements interact in a way that can lead to the emergence or disappearance of specific entanglement transitions. This interplay could potentially be applied to error correction codes based on random circuits (discussed in Secs.~\ref{appcode} and \ref{purification}).

\begin{figure}[hpbt]
\centering
\includegraphics[scale=0.25]{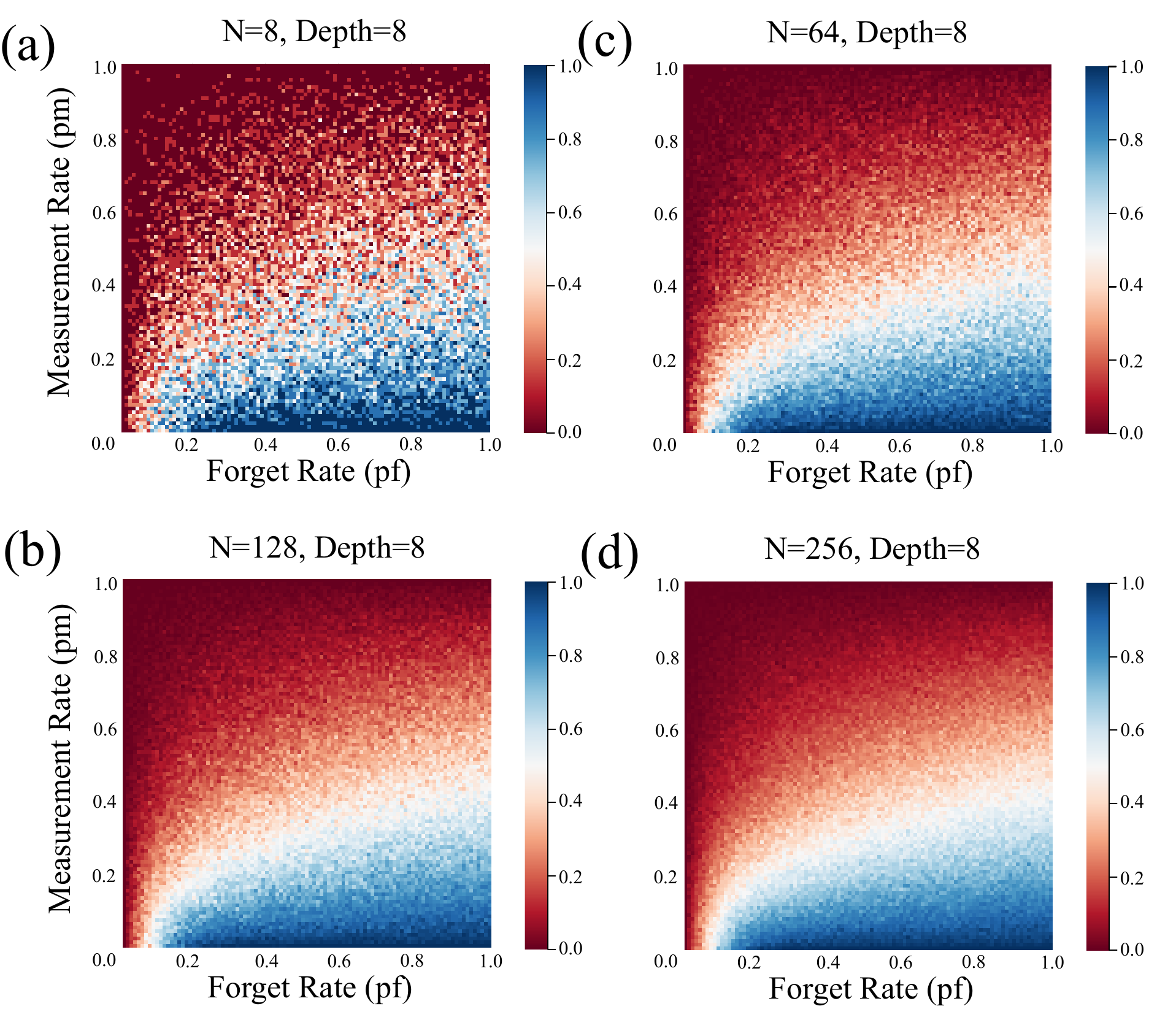}
\caption{The entropy plot under the competition of the measurement and forget processes, where the initial state is pure. The vertical axis is the measurement rate $p_m$ in the circuit. The horizontal axis is the forgetting rate $p_f$. The color bar shows the entropy density $S/N$, with a bluer shade signifying greater entropy. Across (a)--(d), for a constant depth of 8, the structure of the plots remains the same even as the system size $N$ grows. Increasing $N$ imparts a smoother character to the plot, but its fundamental structure persists.}
\label{Fixdepth}
\end{figure}

\begin{figure}[hpbt]
    \centering
    \includegraphics[scale=0.25]{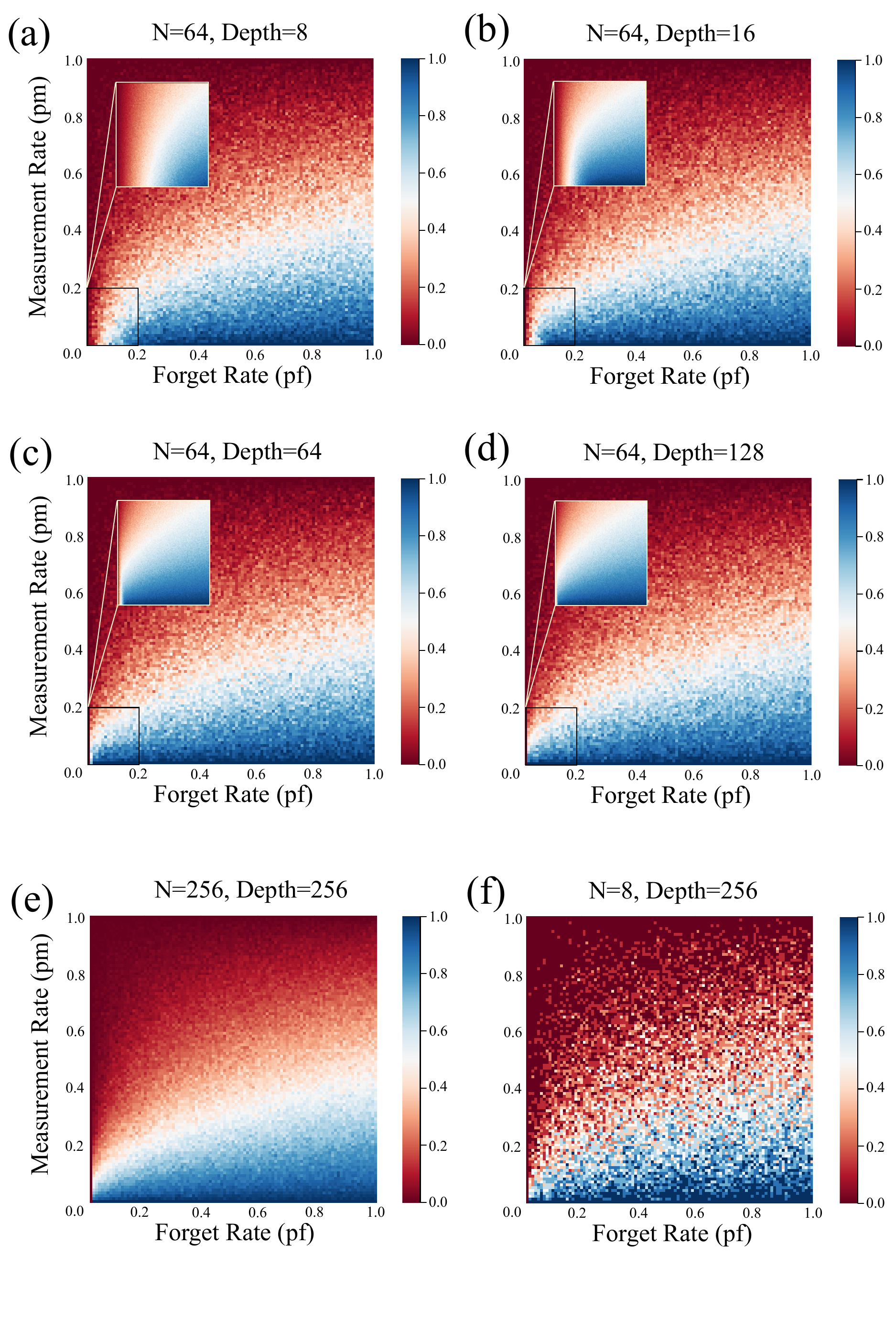}
    \caption{The entropy plot under the competition of the measurement and forget processes, with a pure product initial state. The vertical axis is the measurement rate $p_m$ and the horizontal axis the forgetting rate $p_f$; the color bar shows the entropy density $S/N$ (bluer shade = greater entropy), averaged over $20$ realizations. Panels (a)--(d) fix the system size $N=64$ and vary the circuit depth; the inset in each magnifies the low-rate corner $p_m,p_f\le0.2$ (marked by the box), where the low-/high-entropy boundary sharpens and shifts to smaller $p_f$ as the depth increases. For a fixed $N=64$, the plot remains essentially unchanged once the depth exceeds $N=64$ [(c) and (d) are nearly identical]. Panels (e) and (f) instead fix a large depth ($=256$) and vary the system size ($N=256$ and $N=8$): the structure is unchanged, confirming that the dynamics are also independent of the overall system size.}
    \label{FixN}
\end{figure}

Figs.~\ref{Fixdepth} and \ref{FixN} show the entropy-density diagram $S(\rho)/N$ in the $(p_m,p_f)$ plane. For random unitaries with local depolarizing noise but \emph{without} recorded measurements, the noise acts as a symmetry-breaking field that drives the system to the maximal-entropy fixed point~\cite{ZhiLi}. When recorded measurements are also present, however, the global entropy density cannot reach its maximum: a fraction $\sim p_m$ of the qubits are projected (and hence left pure) in the most recent layer, so on average $\langle S(\rho)/N\rangle \lesssim 1-p_m$. The diagram therefore separates into a low-entropy, measurement-dominated region and a high-entropy, forgetting-dominated region, rather than being uniformly thermalized. This structure is consistent with the known steady-state phase diagram of this model~\cite{DiasNoise}; below, we focus on its depth- and size-dependence, which reveal the underlying thermalization dynamics.

Fig.~\ref{Fixdepth} demonstrates that altering system size while maintaining depth (e.g., Depth~$=8$) does not change the system's fundamental structure. While the entropy transfers are smoother with larger $N$, the underlying dynamics remain unaffected. This outcome can be attributed to the local thermalization rate demonstrated in Fig.~\ref{Forget_only}.

Fig.~\ref{FixN} shows how the plot evolves when the system size is fixed ($N=64$) and the depth is varied. As the depth increases, the high-entropy (forgetting-dominated) region grows and the boundary separating it from the low-entropy (measurement-dominated) region moves to smaller forgetting rates. This change is concentrated in the low-rate corner, which is magnified in the inset of each of panels (a)--(d): the boundary visibly sharpens and shifts toward $p_f\to0$ as the depth grows. Beyond a certain depth, however, the plots stop changing---Fig.~\ref{FixN}(c) and Fig.~\ref{FixN}(d) (depths $64$ and $128$) are nearly identical---indicating that the structure converges once the depth exceeds $\sim N$. The converged plot closely resembles the $N=256$, Depth~$=256$ case [Fig.~\ref{FixN}(e)] and the $N=8$, Depth~$=256$ case [Fig.~\ref{FixN}(f)], confirming that the structure is independent of the overall system size.

This depth-dependence is simply the thermalization dynamics of Sec.~\ref{forgetinduced} resolved in the $(p_m,p_f)$ plane: the plot stops evolving once the depth exceeds $\sim N$, and the converged ($d\gtrsim N$) plot reproduces the steady-state limit.
\\

\section{Implications for Decoding under Partial Information Loss}
\label{appcode}

In this section, we will consider how measure-and-forget dynamics might be applied to Yoshida's decoding protocol. Previous work \cite{YoshidaCode} studied the creation of entanglement in a random quantum circuit with measurement, and showed that there is a duality between the problems of distilling entanglement between two disjoint subsystems and decoding a dynamically evolving classical error-correcting code. We can think of the initial information being encoded in a set of Pauli operators, which the scrambling dynamics of a random quantum circuit causes to evolve and spread. The problem of decoding classical information stored in this code is dual to the problem of distilling entanglement between disjoint subsystems in the random circuit. The measurements at random locations in the circuit correspond on the one hand to syndrome measurements (for the decoding problem) and to steps in the distillation process (for the distillation problem).

The relationship between recoverability of the stored information and the presence of entanglement between $A$ and $B$ can be summarized in the following theorem from \cite{YoshidaCode}:
\\

\textbf{Theorem 1:} In a monitored Clifford circuit, a subsystem $A$ is maximally entangled with its complement $B$, with their mutual information $I_{(A:B)}=2n_A$ ($n_A$ is the number of qubits in the subsystem $A$), if and only if the initial information in the dual classical error-correcting code is fully recoverable \cite{YoshidaCode}.
\\

From the above, recoverability is controlled by the mutual information: if $I(A{:}B)$ is close to its maximum $2n_A$, the stored information can be recovered, and vice versa. More precisely, $I(A{:}B)$ is a graded measure of the recoverable correlation: full recovery requires the saturated value $I=2n_A$ (Theorem 1), whereas an unsaturated $0<I<2n_A$ signals only partial recoverability. In the finite-depth regime studied below, $I(A{:}B)$ generally remains well below $2n_A$, so we are characterizing how much recoverable correlation survives---and how it scales---rather than claiming full recovery. Once forgetting is present the global state is mixed, so the relevant diagnostic is $I(A{:}B)$ rather than the bipartite entanglement entropy of either subsystem---a distinction we rely on throughout this section. In particular, the entropy plots of Sec.~\ref{competition} track the total mixedness and do not by themselves establish that recoverable correlations persist.

Yoshida's original protocol requires a perfect record of measurement outcomes and cannot tolerate any forgetfulness \cite{YoshidaCode}, but he speculated that a robust encoding might tolerate some loss under a modified scheme. Our results bear on this possibility. Using the same monitored circuit for both the distillation and decoding problems \cite{YoshidaCode}, we ask quantitatively how large a forgetting rate can be sustained while $I(A{:}B)$ remains appreciable, and how the correlation scales with system size and depth.

\begin{figure}[hpbt]
    \centering
    \includegraphics[width=0.95\columnwidth]{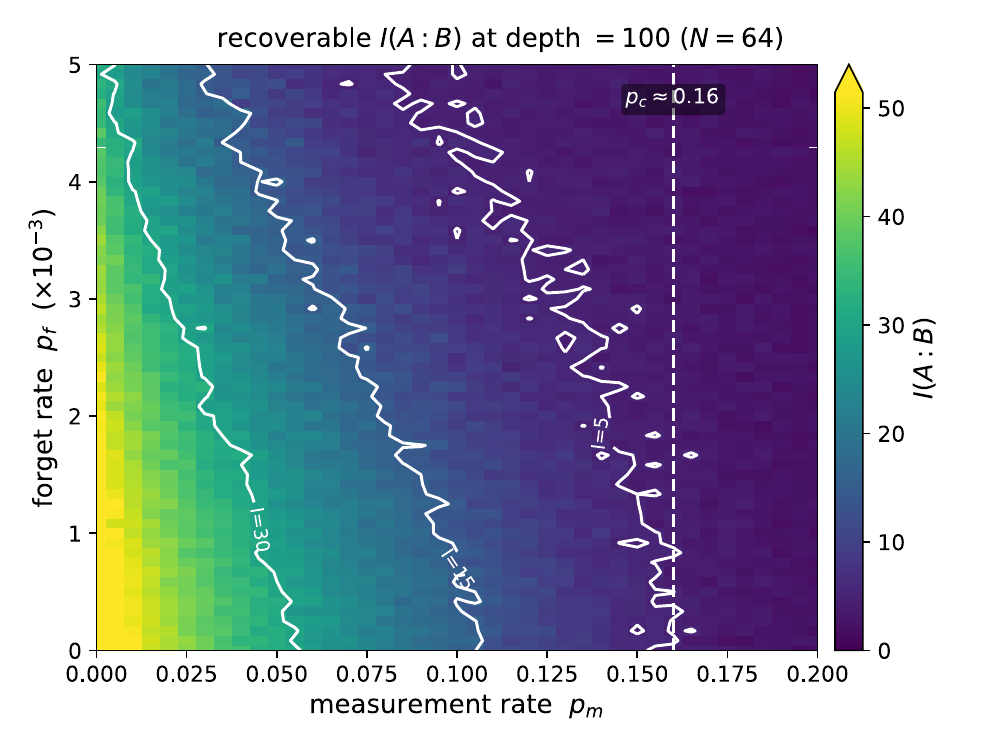}
    \caption{Mutual information $I(A{:}B)$ in the $(p_m,p_f)$ plane at $N=64$, depth $=100$, starting from a pure product state. (The depth is chosen near the mutual-information peak; at depth $256$ the signal has already decayed.) White contours mark $I=30,15,5$; the dashed line is the asymptotic measurement-induced critical point $p_c\approx0.16$. The mutual information is large only at low $p_m$ \emph{and} low $p_f$ and collapses well before $p_c$, since at finite depth the mutual information crossover precedes the asymptotic transition. Each cell is averaged over $16$ realizations.}
    \label{heatmap}
\end{figure}

To quantify how forgetting degrades recoverability, we computed the mutual information $I(A{:}B)$ between the two halves of the chain, using the mixed-state expression $I=S_A+S_B-S_{AB}$ with $S_{AB}=N-\mathrm{rank}$. Two features emerge. First, in the $(p_m,p_f)$ plane shown in Fig.~\ref{heatmap}, at a depth near the mutual-information peak, the mutual information is appreciable only in a small corner of low $p_m$ \emph{and} low $p_f$: it collapses across the measurement-induced transition along $p_m$ and is suppressed by forgetting along $p_f$. Thus, $I(A{:}B)$ is, in fact, highly sensitive to the measurement rate and recoverability survives only in a narrow window.

\begin{figure*}[hpbt]
    \centering
    \includegraphics[width=0.95\textwidth]{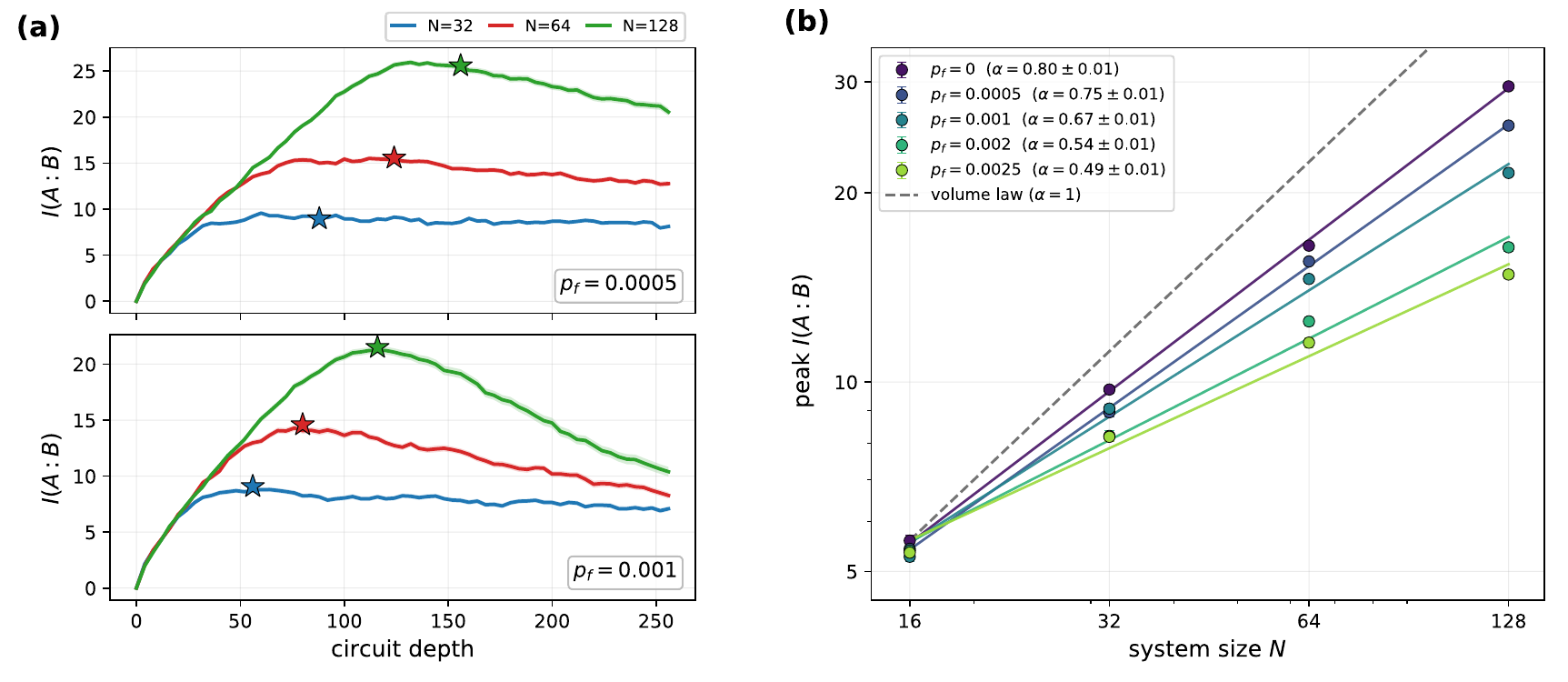}
    \caption{Forgetting drives the mutual information toward sub-extensive scaling. (a) $I(A{:}B)$ versus circuit depth for $p_f=0.0005$ (top) and $p_f=0.001$ (bottom) at $p_m=0.1$, for $N=32,64,128$, starting from a pure product state. The shading is $\pm$ the standard error of the mean (SEM) and stars mark the peaks. Each curve rises to a peak that is higher and later for larger $N$, a finite-depth transient. (b) Peak height $I_{\mathrm{peak}}$ versus $N$ (log--log) for several $p_f$; solid lines are fixed-slope fits with exponent $\alpha(p_f)$, the dashed line is the volume law $\alpha=1$. The effective exponent decreases from $\alpha\approx0.80$ ($p_f=0$) to $\approx0.49$ ($p_f=0.0025$), always below $1$. Here $p_f=0$ is under $p_m=0.1$; a no-measurement, no-forgetting control gives $\alpha\approx1.0$. The statistical error bars are the SEM, and the number of realizations is $=200$; the dominant uncertainty in $\alpha$ is a finite-size systematic error ($\sim\pm0.05$).}
    \label{pillar2}
\end{figure*}

Second, for fixed small $p_f$ the mutual information does not saturate but develops a finite-depth peak that rises to a maximum and then decays (Fig.~\ref{pillar2}(a)); the peak grows higher and occurs later as $N$ increases. Comparing the peak height across system sizes (Fig.~\ref{pillar2}(b)), $I_{\mathrm{peak}}\sim N^{\alpha(p_f)}$ with an effective exponent that decreases monotonically from $\alpha\approx0.80$ at $p_f=0$ (with $p_m=0.1$) to $\alpha\approx0.49$ at $p_f=0.0025$, in every case robustly below the volume-law value $\alpha=1$. A control case with no measurement or forgetting reproduces $\alpha\approx1.0$, confirming that the suppression is a genuine effect of the measurement and forgetting, not a numerical artifact of how we extract the peak. We stress that $\alpha(p_f)$ is an effective finite-size (crossover) exponent; the systematic uncertainty, estimated by dropping the smallest size, is $\sim\pm0.05$. The continuous variation of $\alpha$ with $p_f$, together with this residual curvature, is the quantitative fingerprint of a crossover rather than a sharp transition. Operationally, forgetting drives the recoverable correlation (as measured by $I(A{:}B)$) from near-volume-law scaling toward area-law scaling: a little forgetting can be tolerated only in the narrow, finite-depth window identified above, reminiscent of the depth--noise trade-off proven for noisy encoders~\cite{NoisyEncoding}. Because the peak is a finite-depth transient, this growth with $N$ should not be extrapolated to the thermodynamic limit: at fixed depth, the mutual information increases with $N$ only up to a certain depth-determined scale, beyond which it saturates and eventually decays [cf.\ Fig.~\ref{pillar2}(a)], so $\alpha(p_f)$ is an effective exponent at finite depth and finite size.

\section{Disappearance of Purification Transition and Error-protected Subspace}
\label{purification}

\begin{figure}[hpbt]
    \centering
    \includegraphics[scale=0.25]{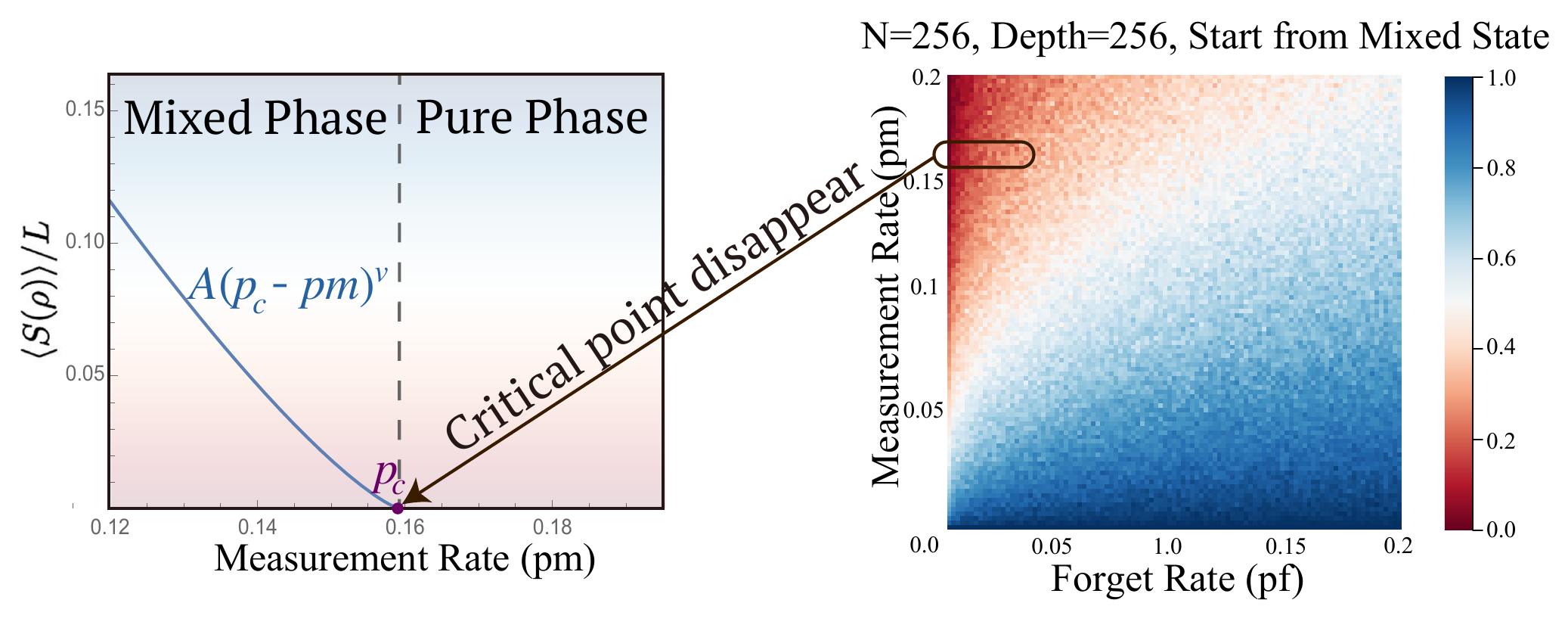}
    \caption{The disappearance of the purification transition. The plot on the left shows the purification transition induced by the rate of random measurement. The critical point $p_c=0.159$ between the mixed phase and the pure phase shows up as the measurement rate $p_m$ approaches $p_c$. Finite-size scaling gives a decay of the form $A(p_c-p_m)^v$ in the large-N limit, where $A\approx7.3$, $\nu\approx1.28$. However, for the plot on the right, when we begin to add the forgetting process, the purification transition is washed out. Here we choose $N=256$, Depth~$=256$, and the initial state as a maximally mixed state.}
    \label{Disappearance}
\end{figure}

In this section, we identify the disappearance of the purification transition when forgetting occurs. By continuously measuring the quantum many-body system, the entropy of its reduced density matrix decreases, resulting in purification. Gullans et al. \cite{MGullans} demonstrate that a balanced interplay between measurements and entangling interactions can lead to a dynamic purification phase transition when the system starts in a mixed state. This transition comprises two phases: (i) a ``pure'' phase, where local purification occurs at a constant rate independent of the system size; and (ii) a ``mixed'' phase where the purification time exponentially grows with the system size.

In the measurement-only setting, the residual entropy density of the mixed phase reflects quantum information that has been encoded into and protected by the dynamics, signaling an error-protected subspace. Within this subspace, quantum information remains reliably encoded and robust against the projective measurements. These codes are of potential significance for fault-tolerant quantum computation due to their high degeneracy and ability to achieve optimal trade-offs between encoded information densities and error thresholds \cite{MGullans}.

We caution, however, that a nonzero residual entropy density does not by itself imply protected quantum information: forgetting also raises the entropy of the global state, but does so through decoherence rather than by encoding logical information. The two cases must therefore be distinguished: residual entropy from measurement-induced encoding (recoverable) versus residual entropy from forgetting (mere mixedness). It is precisely the recoverable, coherent part that is destroyed when forgetting is switched on, even as the total mixedness increases.

In Fig.~\ref{Disappearance}, we see that if the measurement results are partially forgotten, the critical point will vanish correspondingly---consistent with the smearing of measurement-induced transitions by bulk noise~\cite{DiasNoise} (see also Ref.~\cite{Paviglianiti2025Breakdown} for related work on information loss in measurement-induced transitions). The blue curve shown in the left plot stands for the purification transition, where the critical point appears at $p_c=0.1593$. For the plot on the right, we studied $N=256$, Depth~$=256$ case, with the initial state a maximally mixed state. From the plot we see that the critical point disappears when the forgetting rate increases. This phenomenon can destroy the error-protected subspace, hence influencing the code properties.

These findings clarify the interplay between measurement, forgetting, and the stability of the error-protected subspace, and indicate that unrecorded measurements place a limit on how robustly such dynamically generated codes can store quantum information.

\section{Conclusion}
\label{conclusion}

In this paper we studied how a ``forget'' (unrecorded-measurement) process reshapes the dynamics of random Clifford circuits, complementing the steady-state picture of this model by focusing on the finite-depth regime. In the forget-only limit we found a size-independent thermalization rate and a power-law dependence of the thermalization turning point on circuit depth, with an effective exponent $v\approx-1.2$ that collapses across system sizes and falls within the $\propto 1/\mathrm{noise}$ critical-depth family established for related noisy-circuit quantities~\cite{CMISpreading,NoisyEncoding}. In the competition between measurement and forgetting, we showed that the global entropy density is on average bounded by $1-p_m$, that its value depends on depth but not on system size, and that the mutual information develops a finite-depth peak whose height scales sub-extensively, $I_{\mathrm{peak}}\sim N^{\alpha(p_f)}$ with $\alpha$ decreasing from $\approx0.8$ to $\approx0.5$ as forgetting increases. Finally, we showed that forgetting destroys the measurement-induced purification transition, removing the residual-entropy signature of an error-protected subspace.

These results suggest several follow-up directions. It would be valuable to derive the turning-point exponent analytically and to understand how the measurement/forgetting competition modifies the noise-only scaling $\propto 1/\mathrm{noise}$. (Our data hint that the peak depth scales more weakly than $1/p_f$.) A controlled, protocol-level study of Yoshida's distillation under a finite rate of unrecorded syndromes---rather than the mutual-information proxy used here---would turn the present recoverability bounds into a concrete statement about fault tolerance.

\section{Acknowledgements}
\label{acknowledgements}
We thank Beni Yoshida, John Preskill, Yimu Bao, Yaodong Li, Ruihua Fan, Zihan Xia and Yi-Zhuang You, for helpful discussions. This work was supported in part by NSF Grant PHYS-2310794 and NSF Grant FET-2316713.

\textit{Code availability.}---The simulations were performed with the open-source {\sc QuantumClifford.jl} package~\cite{QuantumCliffordJl}; the scripts used to generate the data and figures in this paper are available from the authors upon reasonable request.


\nocite{*}
\bibliography{reference.bib}

\end{document}